\documentclass{article}

\usepackage{arxiv}

\usepackage[utf8]{inputenc} 
\usepackage[T1]{fontenc}    
\usepackage{hyperref}       
\usepackage{url}            
\usepackage{booktabs}       
\usepackage{amsfonts}       
\usepackage{nicefrac}       
\usepackage{microtype}      
\usepackage{graphicx}
\usepackage{doi}

\usepackage{bm} 
\usepackage{amsbsy} 
\usepackage{float}

\usepackage{threeparttable} 

\title{Sigma Basis sets: a new family of GTO basis sets for molecular calculations}

\author{ \href{https://orcid.org/0000-0002-3003-213X}{\includegraphics[scale=0.06]{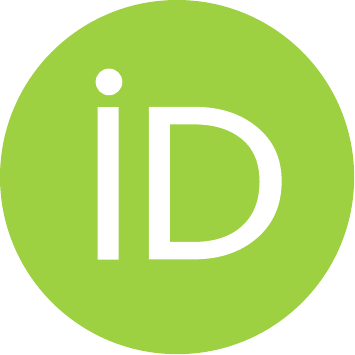}\hspace{1mm}Ignacio Ema}\thanks{
        } \\
    Departamento de Química Física Aplicada, \\
    Facultad de Ciencias, \\
    Universidad Autónoma de Madrid, \\
    Spain \\
    \texttt{nacho.ema@uam.es} \\
    \And
    \href{https://orcid.org/0000-0002-6957-1050}{\includegraphics[scale=0.06]{orcid.pdf}\hspace{1mm}Guillermo Ramírez} \\
    Departamento de Química Física Aplicada, \\
    Facultad de Ciencias, \\
    Universidad Autónoma de Madrid, \\
    Spain \\
    \texttt{guillermo.ramirez@uam.es} \\
    \And
    \href{https://orcid.org/0000-0002-9676-7534}{\includegraphics[scale=0.06]{orcid.pdf}\hspace{1mm}Rafael López} \\
    Departamento de Química Física Aplicada, \\
    Facultad de Ciencias, \\
    Universidad Autónoma de Madrid, \\
    Spain \\
    \texttt{rafael.lopez@uam.es} \\
    \And
    \href{https://orcid.org/0000-0002-1940-422X}{\includegraphics[scale=0.06]{orcid.pdf}\hspace{1mm}José Manuel García de la Vega} \\
    Departamento de Química Física Aplicada, \\
    Facultad de Ciencias, \\
    Universidad Autónoma de Madrid, \\
    Spain \\
    \texttt{garcia.delavega@uam.es} \\
}

\renewcommand{\shorttitle}{\textit{arXiv} Sigma basis sets}
\hypersetup{
pdftitle={A template for the arxiv style},
pdfsubject={q-bio.NC, q-bio.QM},
pdfauthor={David S.~Hippocampus, Elias D.~Striatum},
pdfkeywords={Gaussian functions, Basis sets, Atomic energies, Molecular calculations},
}

\begin{document}

\newcommand{\be}{\begin{equation}}
\newcommand{\ee}{\end{equation}}
\newcommand{\ba}{\begin{array}}
\newcommand{\ea}{\end{array}}
\newcommand{\baa}{\begin{eqnarray}}
\newcommand{\eaa}{\end{eqnarray}}
\newcommand{\Dcal}{{\mathcal{D}}}
\newcommand{\Mcal}{{\mathcal{M}}}
\newcommand{\Ncal}{{\mathcal{N}}}
\newcommand{\Rcal}{\mathcal{R}}  
\newcommand{\Vcal}{\mathcal{V}}
\newcommand{\Ane}{\mathbf{A}}
\newcommand{\Bne}{\mathbf{B}}
\newcommand{\Cne}{\mathbf{C}}
\newcommand{\Dne}{\mathbf{D}}
\newcommand{\Ene}{{\bf E}}
\newcommand{\Fne}{{\bf F}}
\newcommand{\kne}{\mathbf{k}}
\newcommand{\Mne}{{\bf M}}
\newcommand{\Pne}{{\bf P}}
\newcommand{\rne}{{\bf r}}
\newcommand{\Rne}{{\bf R}}
\newcommand{\Sne}{{\bf S}}
\newcommand{\Pop}{\hat{\mathcal{P}}}
\newcommand{\sgbs}{$\sigma$BS}
\newcommand{\sgbshf}{$\sigma$BSHF}
\newcommand{\sgdz}{$\sigma$DZ}
\newcommand{\sgtz}{$\sigma$TZ}
\newcommand{\sgnz}{$\sigma$NZ}
\newcommand{\sgdzhf}{$\sigma$DZHF}
\newcommand{\sgtzhf}{$\sigma$TZHF}
\newcommand{\sgnzhf}{$\sigma$NZHF}
\newcommand{\Gtilde}{{\widetilde{G}}}
\newcommand{\prtd}[1]{\frac{\partial}{\partial #1}}
\newcommand{\prtdd}[2]{\frac{\partial #1}{\partial #2}}
\newcommand{\prtds}[1]{\frac{\partial^2}{\partial #1^2}}
\newcommand{\prtdss}[2]{\frac{\partial^2 #1}{\partial #2^2}}
\newcommand{\prtdsss}[3]{\frac{\partial^2 #1}{\partial #2 \partial #3}}
\newcommand{\prtddss}[2]{\frac{\partial^2}{\partial #1 \partial #2}}
\newcommand{\la}{{\langle \, }}
\newcommand{\ra}{{\, \rangle}}
\newcommand{\bv}{{\, | \,}}

\newcommand{\azul}[1]{{\color{blue}{#1}}}
\newcommand{\verde}[1]{{\color{OliveGreen}{#1}}}
\newcommand{\rojo}[1]{{\color{red}{#1}}}

\def\arraystretch{1.5} 

\maketitle

\begin{abstract}
    A new family of Gaussian-type basis sets named sigma basis sets is presented and preliminarily tested. Sigma basis sets for H, C, N, O and P are reported and their performance is tested in some atomic and molecular calculations.
\end{abstract}


\section{Introduction}

A new family of Gaussian type basis sets (GTO BS), named sigma basis sets ($\sigma$BS) is presented. The basis sets range from DZ to QZ and have the same composition per shell as Dunning basis sets\cite{Dunning1989,Woon1993}, but with different treatment of contractions. 
In particular, two features distinguish $\sigma$BS from currently used BS like Dunning's:
first, if a given primitive appears 
multiplied by a spherical harmonic of quantum number $l = L$, 
all the primitives with the same exponent and $l < L$ are also present in the set, and second, all the BS functions, core, valence and polarization functions, are true contractions, i.e. linear combinations of {\it several} gaussian primitives. 

This work is organized as follows. In Section II, the procedure for developing sigma BS is outlined, and the contraction scheme and composition are described for some selected
atoms of second and third row of the periodic system. In Section III, some preliminary testing on atomic and molecular calculations is reported.

\section{Development of sigma basis sets}
\label{sec:2}

Normalized spherical Gaussian primitive functions (pGTO), $g_{lm}$, are defined 
as:
  
\be \label{eq:1.1}
g_{lm}(\xi,\rne) = \Ncal_{l,\xi}^{r} \; \Ncal_{lm}^\Omega \; 
e^{-\xi \, r^2} \; z_l^m(\rne)
\ee
where $z_l^m (\rne)$ are unnormalized regular solid harmonics:  
  
\be   \label{eq:1.2}  
z_l  (\rne) = r^l \; (-1)   \; P_l^{|m|}(\cos \theta)  
\;   
\left\{  
\begin{array}{lr}  
\cos m \phi & \hspace*{5mm} (m \ge 0) \\  
\sin |m| \phi & \hspace*{5mm}  (m <0)  
\end{array}  
\right.  
\ee  
$P_l^m$ being the associate Legendre functions (see ~\cite{Gradshteyn} Eq 8.751.1), and $\Ncal_{\xi_i}$ and $\Ncal_{LM}^\Omega$, the 
radial and angular normalization constants given by:

\be \label{eq:1.3}
\Ncal_{l,\xi}^{r} = 2^{l+1} \; \sqrt{\frac{\xi^l}{(2l+1)!!}} \; \; 
\left[\frac{(2 \, \xi)^3}{\pi}\right]^{1/4} 
\ee

\be \label{eq:1.4}
\Ncal_{lm}^\Omega = \left[\frac{(2l+1) \; (l-|m|)!}{2 \, (1+\delta_{m0}) \;
\pi \; (l+|m|)!}\right]^{1/2}
\ee

Usual GTO BS consist of contractions 
of Gaussian primitives. They are known as contracted Gaussian functions (CGTO), 
$G_{lm}$, 
defined as:

\be \label{eq:1.5}
G_{lm}(\rne) = \sum_{i=1}^{N_G} c_i \; g_{lm,i}(\rne)
\ee
where $i$ labels the primitives in the set, and the coefficients $c_i$ 
are determined by different procedures 
subject to the 
normalization of the GTO. The number of coefficients
and their values depend on the recipe used in the construction of a particular basis set.

In case of $\sigma$BS, if a primitive with a given exponent $\xi_i$ appears with an angular factor with a quantum number $L$, primitives with the same exponent and all angular factors
with $l < L$ are also present in the $\sigma$BS. Furthermore, all primitives in a given shell
(set of functions with the same angular factor) participate in all 
contractions of the shell. The combination of both features makes it possible to systematically increase the number of primitives in the contractions and, therefore, to improve the quality of the BS functions without penalizing the computational cost. In short, the contractions in $\sigma$BS are built from the same set of exponentials combined with different angular functions. Moreover, while in Dunning BS polarization functions consist of simple Gaussian functions (one primitive per function), in sigma basis sets they are true contractions, leading to significant improvements in energy results.

To simplify the notation, XZ will be used below as an abbreviation for Dunning cc-pVXZ  families. The equivalent $\sigma$BS will be denoted as $\sigma$XZ. In each of the sigma families, we have considered basis sets ranging from X = 2 ($\sigma$DZ) to 4 ($\sigma$QZ).

\begin{table}[H]
\caption{Composition of sigma basis sets}
\centering
\begin{tabular}{|l|c|rl|l|c|rl||rl|}
\hline
Basis  & $N_{exp}^a$ & $\#$ & Primitives &  
Basis  & $N_{exp}^a$ & $\#$ & Primitives & $\#$ & Contractions$^b$ 
\\
\hline \hline
\multicolumn{10}{|c|}{H atom} \\
\hline
\multicolumn{4}{|c|}{Dunning} &
\multicolumn{4}{c||}{Sigma}    & & \\           
\hline
        DZ     &  5 &  7 & ( 4s, 1p)                 &
$\sigma$DZ     & 10 & 19 & (10s, 3p)                 & 5  & [2s, 1p]        \\
        TZ     &  8 & 16 & ( 5s, 2p, 1d)             &
$\sigma$TZ     & 10 & 37 & (10s, 4p, 3d)             & 14 & [3s, 2p, 1d]   \\
        QZ     & 12 & 32 & ( 6s, 3p, 2d, 1f)         &
$\sigma$QZ     & 10 & 66 & ( 4s, 3p, 2d, 1f)         & 30 & [4s, 3p, 2d, 1f]  \\
\hline
\multicolumn{10}{|c|}{C, N and O atoms} \\
\hline
\multicolumn{4}{|c|}{Dunning} &
\multicolumn{4}{c||}{Sigma}    & & \\ 
\hline
        DZ     & 14 & 26 & ( 9s,  4p, 1d)            &
$\sigma$DZ     & 15 & 60 & (15s, 10p, 3d)            & 14 & [3s, 2p, 1d]      \\
        TZ     & 19 & 42 & (10s,  5p, 2d, 1f)        &
$\sigma$TZ     & 15 & 86 & (15s, 10p, 4d, 3f)        & 30 & [4s, 3p, 2d, 1f]  \\
        QZ     & 24 & 68 & (12s,  6p, 3d, 2f, 1g)    &
$\sigma$QZ     & 15 &125 & (15s, 10p, 5d, 4f, 3g)    & 55 & [5s, 4p, 3d, 2f, 1g] \\
\hline
\multicolumn{10}{|c|}{P atom} \\
\hline
\multicolumn{4}{|c|}{Dunning} &
\multicolumn{4}{c||}{Sigma}    & & \\ 
\hline
        DZ     & 21 & 41 & (12s,  8p, 1d)             &
$\sigma$DZ     & 19 & 79 & (19s, 15p, 3d)             & 18 & [4s, 3p,  1d]      \\
        TZ     & 27 & 59 & (15s,  9p, 2d, 1f)         &
$\sigma$TZ     & 19 & 95 & (19s, 15p, 4d, 3f)         & 34 & [5s, 4p,  2d, 1f]  \\
\hline
\hline
\end{tabular}

\vspace*{2mm}
\begin{tablenotes}[flushleft]
\item $^a$ Number of different exponents in the BS.
\item $^b$ Number of contractions per shell is equal for both BS.
\end{tablenotes}
\label{tab:table}
\end{table}

In a previous work\cite{Ema2022}, $\sigma$BS have been reported for He, and their performance
has been tested on the dimer, He$_2$. Here, $\sigma$BS are presented for H, C, N, O and P,
whose composition is detailed in Table 1. The 
numbers of exponentials and primitives are quoted for each BS, as well as the 
number of contractions. Notice that the composition (contractions per shell)
is the same for both families of BS, the only difference being the number of primitives, smaller 
in Dunning BS than in $\sigma$BS. On the other hand, except for the DZ and TZ of H and DZ of C, N and O, the number of exponentials 
is smaller in $\sigma$BS.

The construction of a given $\sigma$BS begins with the selection of a set of exponents to build the primitives of all shells. 
Contraction coefficients are chosen to optimize 
energy of the atomic ground state,
and low lying excited states close to it when applicable, at several computational levels.
In particular, for the atoms selected in this work, except H, core functions are optimized with a Multi-configuration (MC) calculation using the orbitals of the (2s,2p) shell as active space, to yield the best 
average energy for the ground state and the two low-lying excited states in the space spanned by the primitives. 
Polarization functions consist of contractions
sharing primitives with core functions and orthogonal to them. These contractions are
optimized to yield the minimum CISD energy.
The procedure is repeated varying the set of exponents until a 
sufficiently satisfactory result is attained.

\section{Results}

As a preliminary test, we have applied the $\sigma$BS to the calculation of atomic energies and some properties 
in homonuclear diatomics of the selected atoms. 
In Table \ref{tabH}, energies in Hartree for H atom and H$_2$ molecule 
computed with $\sigma$BS are
reported at two computational levels: Hartree-Fock (HF) and CISD. 
The differences, in mE$_h$, between them and the values computed with Dunning
BS, $E(\sigma BS) - E(Dunning)$, appear in parenthesis. The Table  \ref{tabH}
clearly shows the better
performance of $\sigma$BS over Dunning BS, both in the atom and the molecule.

For the remaining selected atoms, energies of the ground state and two 
low-lying excited states are collected in Tables \ref{tabCN} and \ref{tabOP}. 
They have been computed also at two different computational levels, in this case: MC with (2s,2p) shell as active space, and CISD.
The results reported correspond to the energies computed with the $\sigma$ BS,
and the differences, in mE$_h$, between them and the values computed with Dunning
BS. Again, these differences, $E(\sigma BS) - E(Dunning)$, appear within parenthesis. 
As it can be seen in these Tables \ref{tabCN} and \ref{tabOP}, 
$\sigma$BS overcome the results of Dunning BS in all cases,
and the improvement increases when passing from the lower to the higher computational level.
The differences are more apparent for the DZ BS, where the MC energies obtained
with the $\sigma$DZ are better than those of Dunning's TZ, and even QZ in some cases.
It is also noticeable that this happens for the three states considered.

\begin{table}[H]
\caption{
Energies in E$_h$ of H and H$_2$(1.4 au) computed with $\sigma$ basis sets. In parenthesis:
$E(\sigma BS) - E(Dunning)$ in mE$_h$.\label{tabH}
}
\begin{center}
\begin{tabular}{|l|c|cc|}
\hline
  & H & \multicolumn{2}{|c|}{H$_2$(1.4 au)}  \\
\hline
 & HF & HF & CISD \\
\hline
$\sigma$DZ   & -0.499995 (-0.717) & -1.133298 (-4.588) & -1.169945 (-6.547)  \\ 
$\sigma$TZ   & -0.499997 (-0.188) & -1.133548 (-0.587) & -1.173265 (-0.930)  \\ 
$\sigma$QZ   & -0.499999 (-0.053) & -1.133619 (-0.160) & -1.174050 (-0.255)  \\ 
$\sigma$5Z   & -0.499999 (-0.004) & -1.133620 (-0.014) & -1.174258 (-0.036)  \\ 
\hline
\end{tabular} 
\end{center}

\end{table}

\begin{table}[H]
\caption{
Energies in E$_h$ for the atomic ground state and low lying excited states of C and N computed with $\sigma$ basis sets. In parenthesis:
$E(\sigma BS) - E(Dunning)$ in mE$_h$.
\label{tabCN} }
\begin{tabular}{|c|c|cc|c|cc|}
\hline
 & \multicolumn{3}{|c|}{C} 
 & \multicolumn{3}{|c|}{N} \\
 \hline
BS & State & MC & CISD  & State & MC & CISD  \\
\hline
            & $^3$P & -37.705833 ( -5.589) & -37.775227 (-10.927) & $^4$S & -54.400495 (-12.081) & -54.499024 (-21.982) \\ 
 $\sigma$DZ & $^1$D & -37.648090 ( -5.396) & -37.722370 (-11.256) & $^2$D & -54.296078 (-13.572) & -54.400815 (-24.110) \\ 
            & $^1$S & -37.610245 ( -5.413) & -37.670705 ( -9.586) & $^2$P & -54.262724 (-14.215) & -54.357170 (-23.753) \\
\hline
            & $^3$P & -37.705593 ( -1.821) & -37.784070 ( -3.672) & $^4$S & -54.400768 ( -3.411) & -54.517425 ( -6.467) \\ 
 $\sigma$TZ & $^1$D & -37.648106 ( -1.789) & -37.736501 ( -4.007) & $^2$D & -54.296120 ( -3.835) & -54.426015 ( -7.816) \\ 
            & $^1$S & -37.610304 ( -1.764) & -37.684698 ( -3.994) & $^2$P & -54.262858 ( -4.005) & -54.382299 ( -7.689) \\
\hline
            & $^3$P & -37.705943 ( -0.332) & -37.786144 ( -0.921) & $^4$S & -54.400835 ( -0.659) & -54.522036 ( -1.631)  \\ 
 $\sigma$QZ & $^1$D & -37.648107 ( -0.332) & -37.739420 ( -1.037) & $^2$D & -54.296122 ( -0.812) & -54.432420 ( -2.076)  \\ 
            & $^1$S & -37.610314 ( -0.314) & -37.687457 ( -1.096) & $^2$P & -54.262899 ( -0.859) & -54.389223 ( -2.096)  \\
\hline
\end{tabular}
\end{table}

\begin{table}[H]
\caption{
Energies in E$_h$ for the atomic ground state and low lying excited states of O and P computed with $\sigma$ basis sets. In parenthesis:
$E(\sigma BS) - E(Dunning)$ in mE$_h$.
\label{tabOP} }
\begin{tabular}{|c|c|cc|c|cc|}
\hline
 & \multicolumn{3}{|c|}{O} 
 & \multicolumn{3}{|c|}{P} \\
 \hline
BS & State & MC & CISD  & State & MC & CISD  \\
\hline
            & $^3$P & -74.809120 (-22.932) & -74.944572 (-37.535) & $^4$S & -340.718417 ( -9.403) & -340.805439 (-15.988) \\ 
 $\sigma$DZ & $^1$D & -74.729188 (-23.827) & -74.867552 (-39.221) & $^2$D & -340.648773 (-10.611) & -340.741252 (-17.550) \\ 
            & $^1$S & -74.664402 (-24.748) & -74.788724 (-38.719) & $^2$P & -340.624395 (-11.132) & -340.708470 (-17.961) \\
\hline
            & $^3$P & -74.809256 ( -6.177) & -74.979149 (-11.681) & $^4$S & -340.718647 ( -2.663) & -340.819650 ( -4.187) \\ 
 $\sigma$TZ & $^1$D & -74.729198 ( -6.450) & -74.905380 (-12.626) & $^2$D & -340.648793 ( -2.966) & -340.763864 ( -5.322) \\ 
            & $^1$S & -74.664513 ( -6.703) & -74.824699 (-12.433) & $^2$P & -340.624529 ( -3.107) & -340.730984 ( -5.472) \\
\hline
            & $^3$P & -74.809273 ( -1.322) & -74.989034 ( -3.199) & $^4$S & $--$ & $--$ \\ 
 $\sigma$QZ & $^1$D & -74.729202 ( -1.421) & -74.916372 ( -3.570) & $^2$D & $--$ & $--$ \\ 
            & $^1$S & -74.664548 ( -1.482) & -74.836291 ( -3.690) & $^2$P & $--$ & $--$ \\
\hline
\end{tabular}
\end{table}

We have also studied the 
energy and internuclear distance in the minimum of the curves
computed at MC (HF for H$_2$) and CISD levels. In Table \ref{diatomics} the results obtained with the $\sigma$BS are compared with those computed with Dunning BS.
The results of Dunning 5Z (QZ for Phosphorus) are taken as a reference to test the quality of the results.

Table \ref{diatomics} shows the excellent performance of the $\sigma$BS, even more neat in small $\sigma$DZ whose results clearly overcome those of Dunning equivalent DZ BS.

\begin{table}[H]
\caption{Ground state minimum energies and equilibrium  distances in homonuclear diatomics
\label{diatomics}}

\begin{tabular}{|c|cc|cc|cc|cc|cc|rr|}
\hline
 & \multicolumn{2}{|c|}{E(MC or HF) / E$_h$}  
 & \multicolumn{2}{|c|}{E(CISD) / E$_h$} 
 & \multicolumn{2}{|c|}{E$_{corr}$ / E$_h$} 
 & \multicolumn{2}{|c|}{R$_e$(HF) / \AA}  
 & \multicolumn{2}{|c|}{R$_e$(CISD) / \AA} \\
\hline
BS &  $\sigma$ & Dun
   &  $\sigma$ & Dun
   & $\sigma$    & Dun
   & $\sigma$   & Dun
   & $\sigma$   & Dun \\
\hline
\multicolumn{11}{|c|}{H$_2$} \\
\hline
 DZ & -1.133328 &  -1.128746 &  -1.169948 & -1.163673 &  -0.036620 & -0.034926 &  0.734367 & 0.747954 & 0.742709 & 0.760922 \\
 TZ & -1.133583 &  -1.132990 & -1.173266 & -1.172337 &  -0.039683 & -0.039347 &  0.733793 & 0.733743 & 0.741868 & 0.742653  \\
 QZ & -1.133657 &  -1.133495 & -1.174050 & -1.173797 &  -0.040394 & -0.040302 &  0.733537 & 0.733743 & 0.741494 & 0.741891  \\
 5Z &           &  -1.133646 &            & -1.174223 &            & -0.040578 &          & 0.733574 &           & 0.741566 \\
\hline
\multicolumn{11}{|c|}{C$_2$} \\
\hline
 DZ & -75.399026 & -75.387049 & -75.747612 & -75.727617 & -0.348587 & -0.339068 & 1.246061 & 1.252332 & 1.265416 & 1.272121 \\
 TZ & -75.404566 & -75.401450 &  -75.784667 & -75.777984 & -0.380101 & -0.376534 & 1.241324 & 1.240889 & 1.252033 & 1.252039 \\
 QZ & -75.405925 & -75.405787 &  -75.795789 & -75.794412 &   -0.389863 & -0.388628  & 1.239207 & 1.238896 & 1.247084 & 1.246993 \\
 5Z & & -75.406545   & & -75.799239  & & -0.392694 &          & 1.238720 &          & 1.245906 \\
\hline
\multicolumn{11}{|c|}{N$_2$} \\
\hline
 DZ & -108.980825 & -108.955559 & -109.310132 & -109.271025 & -0.329307 & -0.315466 &  1.069024 & 1.077301 & 1.110994 & 1.120107 \\
  TZ &  -108.991846 & -108.986557 & -109.372824 & -109.361374 & -0.380977 & -0.374817 &  1.067394 & 1.067112 & 1.104875 & 1.104700 \\
  QZ & -108.994369 & -108.994470 &  -109.391694 & -109.389591 &    -0.397326 & -0.395121 & 1.065819 &  1.065583 &  1.101412 &  1.101203 \\
  5Z & & -108.996190 & & -109.398595 & &  -0.402407 &      & 1.065408 &          & 1.100384 \\
\hline
\multicolumn{11}{|c|}{O$_2$} \\
\hline
  DZ & -149.645122 & -149.613606  & -150.031401 & -149.972402 & -0.386279  & -0.358796 & 1.159063  & 1.154302 &  1.230114  &  1.218938 \\
  TZ & -149.667415 & -149.658421 & -150.121984 & -150.102449 &  -0.454568 & -0.444028 & 1.153749 &  1.152330  &  1.214676  &  1.213689   \\
  QZ & -149.670882 & -149.670126 &   -150.147865 &  -150.142986 & -0.476983 & -0.472821 &   1.152156 &  1.151011 &  1.210668 &  1.209140 \\
  5Z & & -149.673149 & &   -150.156534 & & -0.483385 &           & 1.150707 &          & 1.208077 \\
\hline
\multicolumn{11}{|c|}{P$_2$} \\
\hline
DZ &  -681.489805 & -681.466354 & -681.716205 & -681.687337 & -0.227629 & -0.222181 & 1.863161 & 1.867287 & 1.901927 & 1.905391 \\
TZ &  -681.498015 & -681.490781 & -681.765100 & -681.756121 & -0.267725 & -0.265974 & 1.855853 & 1.857063 & 1.883494 & 1.884618 \\
QZ  &             & -681.498432 &             & -681.777667 &           & -0.279665 &          & 1.852693 &          & 1.875288 \\
\hline
\end{tabular}
\end{table}

All calculations have been carried out with MOLPRO\cite{MOLPRO}.
Ancillary material is supplied in file \texttt{anc/Sigma\_BS\_HCNOP.txt} which contains 
the $\sigma$BS for the selected atoms in MOLPRO's format. 



\end{document}